\documentstyle[12pt,aaspp4]{article}
\lefthead{Athanassoula \& Bureau}
\righthead{Bar Diagnostics in Edge-On Spiral Galaxies}
\slugcomment{Accepted for publication in The Astrophysical Journal}
\begin{document}
%
% Definitions.
%
\newcommand{\kms}{km~s$^{-1}$ }
\newcommand{\msun}{$M_{\mbox{\scriptsize \sun}}$ }
%
% Paper
%
\title{Bar Diagnostics in Edge-On Spiral Galaxies. II. Hydrodynamical
Simulations.}
\author{E.\ Athanassoula}
\affil{Observatoire de Marseille, 2 place Le Verrier, F-13248 Marseille Cedex
4, France}
\and
\author{M.\ Bureau\altaffilmark{1}}
\affil{Mount Stromlo and Siding Spring Observatories, Institute of Advanced
Studies, The Australian National University, Private Bag, Weston Creek P.O.,
ACT~2611, Australia}
\altaffiltext{1}{Now at Sterrewacht Leiden, Postbus~9513, 2300~RA Leiden, The
Netherlands}
\begin{abstract}
We develop diagnostics based on gas kinematics to identify the presence of a
bar in an edge-on spiral galaxy and determine its orientation. We use
position-velocity diagrams (PVDs) obtained by projecting edge-on
two-dimensional hydrodynamical simulations of the gas flow in a barred galaxy
potential. We show that when a nuclear spiral is formed, the presence of a gap
in the PVDs, between the signature of the nuclear spiral and that of the outer
parts of the disk, reliably indicates the presence of a bar. This gap is due
to the presence of shocks and inflows in the simulations, leading to a
depletion of the gas in the outer bar region. If no nuclear spiral signature
is present in a PVD, only indirect arguments can be used to argue for the
presence of a bar. The shape of the signature of the nuclear spiral, and to a
lesser extent that of the outer bar region, allows to determine the
orientation of the bar with respect to the line-of-sight. The presence of dust
can also help to discriminate between viewing angles on either side of the
bar. Simulations covering a large fraction of parameter space constrain the
bar properties and mass distribution of observed galaxies. The strongest
constraint comes from the presence or absence of the signature of a nuclear
spiral in the PVD.
\end{abstract}
\section{Introduction\label{sec:introduction}}
\nopagebreak
The importance of bars in the structure of spiral galaxies is recognised in
the Hubble sequence for the classification of galaxies (Sandage
\markcite{s61}1961). In the first paper of this series (Bureau \& Athanassoula
\markcite{ba99}1999, hereafter Paper~I), we highlighted the difficulties
involved in the identification of bars in edge-on systems. It is clear that
the detection of such bars based on photometric or morphological criteria
(e.g.\ de Carvalho \& da Costa \markcite{dd87}1987; Hamabe \& Wakamatsu
\markcite{hw89}1989) is uncertain. Kuijken \& Merrifield (\markcite{km95}1995,
hereafter KM95; see also Merrifield \markcite{m96}1996) were the first to show
that the particular kinematics of barred disks could be used to identify bars
in edge-on spirals. They showed that the periodic orbits in an edge-on barred
galaxy produce characteristic double-peaked line-of-sight velocity
distributions, which can be taken as the signature of a bar.

In \markcite{ba99}Paper~I, we improved on the work of \markcite{km95}KM95. We
studied the signatures of individual periodic orbit {\em families} in the
position-velocity diagrams (PVDs) of edge-on barred spirals (before combining
them to model real galaxies), and examined how the PVDs depend on the viewing
angle. We adopted a widely used mass model and a well-defined method to
populate the periodic orbits. Our aim was to provide insight into the
projected kinematical structure of barred disks, and to provide guidance to
interpret stellar and gaseous kinematical observations of edge-on spiral
galaxies. We showed in \markcite{ba99}Paper~I that the global appearance of a
PVD can be used as a diagnostic to detect the presence of a bar in an edge-on
disk. The signatures of the various periodic orbit families leave gaps in the
PVDs which are a direct consequence of the non-homogeneous distribution of
orbits in a barred spiral. The signature of the $x_1$ periodic orbits is
parallelogram-shaped and occupies all four quadrants of the PVDs, reaching
very high radial velocities when the bar is seen end-on and only low
velocities when the bar is seen side-on. The signature of the $x_2$ orbits,
when present, is similar to that of the $x_1$, but reaches its maximum radial
velocities at opposite orientations. Those features can be used to
determine the 
viewing angle with respect to the bar in an edge-on disk. However, even if
carefully chosen and populated, periodic orbits provide only an approximation
to the structure and kinematics of the stars and gas in spiral galaxies. For
example, a number of stars may be on irregular orbits, and shocks can develop
in the gas.

In this paper (Paper~II), we concentrate on developing bar diagnostics for
edge-on disks using the gaseous component alone. We use the hydrodynamical
simulations of Athanassoula (\markcite{a92b}1992b, hereafter
\markcite{a92b}A92b), designed to study the gas flow and shock formation in
barred spiral galaxies. Unlike \markcite{ba99}Paper~I, these simulations
properly take into account the fact that the gas is not a collisionless
medium. The shocks and inflows which develop in the simulations lead to better
bar diagnostics than those of \markcite{ba99}Paper~I. In addition, we run
simulations covering a large fraction of the parameter space likely to be
occupied by real galaxies. The PVDs produced can thus be directly compared
with observations not only to detect the presence of bars in edge-on spiral
galaxies, but also to constrain the mass distribution of the systems observed.
In particular, Bureau \& Freeman (\markcite{bf99}1999, hereafter BF99) have
applied those diagnostics to long-slit spectroscopic observations of a large
number of edge-on spiral galaxies, most of which have a boxy or peanut-shaped
bulge, to determine the formation mechanism of these objects and
study the vertical structure of bars. In \markcite{ab99}Paper~III
(Athanassoula \& Bureau \markcite{ab99}1999), using fully self-consistent
three-dimensional (3D) $N$-body simulations, we will develop similar bar
diagnostics for the stellar (collisionless) component of barred spiral
galaxies.

We describe the mass model and hydrodynamical simulations used in this paper
in \S~\ref{sec:hydro}. In \S~\ref{sec:bardiag}, we study the signatures in the
PVDs of the various components present in the simulations and discuss the
influence of the parameters of the mass model. The effects of dust extinction
are illustrated in \S~\ref{sec:dust}. We develop the bar diagnostics for
edge-on disks and discuss the limitations of our
models for the interpretation of real data in
\S~\ref{sec:discussion}. We conclude in 
\S~\ref{sec:conclusions} with a brief summary of our main results.
\section{Hydrodynamical Simulations\label{sec:hydro}}
\nopagebreak
For the hydrodynamical simulations, we use the flux-splitting second-order
scheme of G.\ D.\ van Albada (van Albada \& Roberts \markcite{vr81}1981; van
Albada, van Leer, \& Roberts \markcite{vvr82}1982; van Albada
\markcite{v85}1985). It is the same code as that used by \markcite{a92b}A92b
so we will only briefly review its main properties here.

The simulations are two-dimensional and time-dependent, and the gas is treated
as ideal, isothermal, and non-viscous. The simulations are not self-consistent
and we do not consider the self-gravity of the gas; the flow is calculated
using the potential described in \markcite{ba99}Paper~I (see also Athanassoula
\markcite{a92a}1992a, hereafter \markcite{a92a}A92a). The mass model has two
axisymmetric components, a Kuzmin/Toomre disk (Kuzmin \markcite{k56}1956;
Toomre \markcite{t63}1963) and a bulge-like spherical density
distribution. They combine to yield a flat rotation curve in the outer parts
of the disk. We use a homogeneous ($n=0$) or inhomogeneous ($n=1$) Ferrers
spheroid (Ferrers \markcite{f77}1877) as a third component representing the
bar. Each model is described by four main parameters: the bar axial ratio
$a/b$, the quadrupole moment of the bar $Q_m$ (proportional to the mass of the
bar), the Lagrangian radius $r_L$ (the radius of the Lagrange points $L_1$ and
$L_2$ on the major axis of the bar, approximately inversely proportional to
the bar pattern speed), and the central concentration $\rho_c$. The other
quantities are fixed, including the semi-major axis of the bar $a=5$~kpc. We
refer the reader to \markcite{ba99}Paper~I for a more detailed description of
the mass model. \markcite{a92a}A92a discusses at great length its relevance to
real galaxies. Suffice it to say here that the properties of the mass model
are in excellent agreement with those of early-type barred spirals.

The simulations are started with a massless bar and an additional axisymmetric
component of mass equal to that of the desired final bar. Mass is transfered
from that component to the bar over $10^8$~yr and the simulations are run
until the gas flow is roughly stationary in the frame of reference corotating
with the bar (about 10 bar revolutions). An $80\times160$~cells grid is used
to cover a disk of 16~kpc radius, assuming bisymmetry. The inner half of the
simulations are then regridded to the $80\times160$ grid and the simulations
continued for another 5 bar revolutions using this increased spatial
resolution ($0.1\times0.1$~kpc$^2$ cell size; see van Albada \& Roberts
\markcite{vr81}1981). Star formation and mass loss are modeled crudely. The
gas density is lowered artificially in high density regions and gas is added
uniformly over the grid. This process is governed by the equation
\begin{equation}
\label{eq:sf}
d\rho_g/dt=\alpha\rho_{g,init}^2-\alpha\rho_g^2,
\end{equation}
where $\rho_g$ is the gas density, $\alpha$ is a constant (set to
0.3~\msun$^{-1}$~pc$^2$~Gyr$^{-1}$ in most runs), and
$\rho_{g,init}=1$~\msun~pc$^{-2}$ is the uniform initial gas density. It
therefore takes about 3~Gyr for the gas to be reprocessed. There is no
artificial viscosity in the code.

Our main tool in this paper will be PVDs, representing the projected density
of material in the edge-on disks as a function of line-of-sight velocity and
projected position along the major axis. Since we are only interested in the
bar region, we use only the inner 8~kpc~$\times$~8~kpc region of the
simulations, covered by the high resolution grid. At larger radii, the motion
of the gas can, to first order, be considered as circular, since the force of
the bar decreases steeply with radius. Including the outer regions thus makes
no difference to our PVDs.

As in \markcite{ba99}Paper~I, the models considered are those of
\markcite{a92a}A92a (see her Table~1). We will also use her units:
$10^6$~\msun for masses, kpc for lengths, and \kms for velocities. It is
essential to understand the orbital structure of the models to interpret
properly the results of the simulations. Orbital properties have been
discussed in \markcite{ba99}Paper~I and \markcite{a92a}A92a, but also in
Athanassoula et al.\ \markcite{abmp83}(1983), Papayannopoulos \& Petrou
\markcite{pp83}(1983), Teuben \& Sanders \markcite{ts85}(1985), and
others. Here, we will mainly use \markcite{ba99}Paper~I and
\markcite{a92a}A92a for comparison. We will also draw heavily on the results
of \markcite{a92b}A92b, which used the same set of simulations but for
different purposes. She discussed in detail the gas flow and compared the
results with the properties of real galaxies.
\section{Bar Diagnostics\label{sec:bardiag}}
\nopagebreak
In this section, we will concentrate on understanding the PVDs of two
inhomogeneous bar models which are prototypes of models with and without inner
Lindblad resonances (ILRs). As suggested in \markcite{a92a}A92a, we will
identify the existence and position of the ILRs with the existence and extent
of the $x_2$ periodic orbits. The two models considered are the same as in
\markcite{ba99}Paper~I: model 001 ($n=1$, $a/b=2.5$, $Q_m=4.5\times10^4$,
$r_L=6.0$, $\rho_c=2.4\times10^4$) and model 086 ($n=1$, $a/b=5.0$,
$Q_m=4.5\times10^4$, $r_L=6.0$, $\rho_c=2.4\times10^4$). We will then extend
our results to other models and analyse how the properties of the PVDs vary as
the parameters of the mass model are changed.
\subsection{Model 001 (ILRs)\label{sec:001}}
\nopagebreak
\placefigure{fig:001}

Figure~\ref{fig:001} shows PVDs for model 001, which has ILRs. The figure
shows, for the inner half of the simulation, the face-on surface density of
the gas and PVDs obtained by projecting the simulation edge-on and using various
viewing angles with respect to the bar. Unlike the models of
\markcite{ba99}Paper~I, which are symmetric around both the minor and major
axes of the bar, the present simulations are bisymmetric, so we need to cover
a viewing angle range of 180\arcdeg. The viewing angle $\psi$ is defined to be
0\arcdeg\ for a line-of-sight parallel to the major axis of the bar and
90\arcdeg\ for a line-of-sight perpendicular to it, increasing
counterclockwise in the surface density plots.

\markcite{a92b}A92b showed convincingly that the two strong parallel narrow
segments present in the surface density plot of Figure~\ref{fig:001} represent
offset shocks on the leading sides of the bar, displaying strong density
enhancements and sharp velocity gradients. They can be identified with the
dust lanes observed in barred spiral galaxies. The structures seen at the ends
of the bar and perpendicular to it are also shocks. In the inner bar region is
a very intense two-arm nuclear spiral, connecting with the offset shocks.
There is little gas in the barred region outside the nuclear spiral, which we
will hereafter refer to as the outer bar region.
Beyond the bar the surface density is almost featureless; only a few spiral
arms are seen. \markcite{a92b}A92b showed that, in the outer bar region, the
streamlines have roughly the shape and orientation of the $x_1$
periodic orbits. As 
one moves inward, the streamlines change gradually to the shape and
orientation corresponding to the $x_2$ periodic orbits. This flow pattern
leads to the offset shocks and results in an inflow of gas toward the nuclear
region, accounting for the gas distribution in the bar: low densities in the
outer bar region and high densities in the center. The velocities are small
(in the reference frame corotating with the bar) around the Lagrange points on
the minor axis of the bar and the flow is close to circular outside the bar
region.

The PVDs in Figure~\ref{fig:001} show the existence of three distinct regions:
an inner region, corresponding to the signature of the nuclear spiral; an
intermediate region, corresponding to the signature of the outer parts of the
bar; and an external region, corresponding to the signature of the parts
outside the bar. It is important to notice how low the density of the
intermediate region is, compared to that of the other two regions. This is not
surprising since, as mentioned in \markcite{a92b}A92b and above, most of the
bar region has very low gas density, with the exception of the central part
(the nuclear spiral) and the two shock loci. This will form the basis of the
bar diagnostics which we will develop later on. Let us now examine each of the
three regions separately, by keeping only the gas in the targeted region (and
masking out the gas in the other regions) when calculating the PVDs.
\placefigure{fig:bar}

Figure~\ref{fig:bar} shows the surface density and PVDs of model 001 when
considering the low density outer bar region only. Because the high density
regions have been masked out, and because we are looking only at relatively low
density regions, we see much more structure in the PVDs of
Figure~\ref{fig:bar} than in the corresponding sections of the PVDs of
Figure~\ref{fig:001}.

When compared with Figure~4 of \markcite{ba99}Paper~I, Figure~\ref{fig:bar}
shows, at first glance, many similarities, but also, when scrutinised closer,
a number of differences. This could be expected since the gas streamlines in
that region follow loosely, but far from exactly, the shape and orientation of
the $x_1$ orbits, which are elongated parallel to the bar.

The parallelogram-shaped signature of the $x_1$ orbits in Figure~4 of
\markcite{ba99}Paper~I is again observed here and the ``forbidden'' quadrants
are again populated. This is due to the fact that both the $x_1$ orbits and
the streamlines in the outer bar region are not circular, but elongated. In
the gas, however, the parallelogram shape is also present for $\psi=0\arcdeg$,
while, for this viewing angle, the PVD of the $x_1$ orbits showed a
bow-shaped feature. The reason is 
that, unlike the $x_1$ orbits, the streamlines do not have their major axes
parallel to that of the bar, but rather at an angle of about 20\arcdeg\ to it
(see \markcite{a92b}A92b). Also, at $\psi=90\arcdeg$, the $x_1$ orbits show a
near-linear PVD, while the gas shows an additional faint bow-shaped feature
reaching high velocities near the center. Masking out a larger fraction of the
simulation (not shown), it is possible to show that this extra feature
arises from the very low density areas just inside the offset
shocks, close to the major axis of the bar. In this section of the bar, the
flow differs substantially from the behaviour of the $x_1$ periodic orbits.
The elongated ``hole'' in the center of the PVDs, at intermediate viewing
angles, is present in both sets of PVDs, and is due to the way the $x_1$
family was terminated and the way the gas density was masked out. This causes
in both cases an absence (or low density) of quasi-circular streamlines near
the end of the bar. In the hydrodynamical simulation, there is also a reduced
degree of symmetry with respect to viewing angles of 0\arcdeg\ or 90\arcdeg,
in particular for the angles 67.5\arcdeg\ and 112.5\arcdeg. They are, of
course, identical in the PVDs of \markcite{ba99}Paper~I. This is because the
gas flow is bisymmetric, as opposed to being symmetric with respect to both
the minor and major axes of the bar, as are the $x_1$ orbits (see
Fig.~2 of \markcite{a92b}A92b). 

In Figure~\ref{fig:bar}, we see that the radial velocities are maximum
(compared to the circular velocity) for lines-of-sight almost parallel to the
bar (highest values for $\psi=0\arcdeg$ and 157.5\arcdeg), and decrease as
$\psi$ increases, to reach a minimum when the line-of-sight is roughly along
the bar minor axis (lowest values for $\psi=90\arcdeg$ and 67.5\arcdeg). The
same behaviour was seen for the PVDs of the $x_1$ orbits
(\markcite{ba99}Paper~I), but with a slight shift of the viewing angle, so
that the maximum and minimum occurred at exactly 0\arcdeg\ and 90\arcdeg,
respectively. This was to be expected since, like the $x_1$ orbits, the
streamlines in the outer bar region are elongated parallel to it (with a
small offset). Similarly, the position where the maximum occurs moves out as
the viewing angle increases from $\psi=0\arcdeg$ to $\psi=90\arcdeg$.  This
was also found for the $x_1$ orbits (\markcite{ba99}Paper~I), where it was
explained by considering the trace of an elongated orbit in a PVD as a
function of viewing angle and distance from the center. The explanation is
analogous here. However, in the hydrodynamical simulation, the maxima
generally occur at a larger distance from the center than in the periodic
orbits approach (this is easily seen by comparing Fig.~\ref{fig:bar} to Fig.~4
in \markcite{ba99}Paper~I). This is due to the fact that the $x_1$-like flow
does not extend up to the center of the simulated galaxy (contrary to the
$x_1$ orbits in \markcite{ba99}Paper~I), but is superseded by an $x_2$-like
flow at a certain radius. For example, the nuclear spiral (which has an
$x_2$-like behaviour) extends about 1.6~kpc on the minor axis of the bar,
which is the projected distance at which the envelope of the signature
of the outer bar region in the PVD for $\psi=0\arcdeg$ drops abruptly.

To see which features of the gas distribution contribute to the PVDs, we have
repeated the blanking exercise, this time masking out all areas except either 
the shock loci along the leading sides of the 
bar or the density enhancements near the ends of the bar major axis (not
shown). The gas in the shock loci does not contribute any outstanding features
to the PVDs, mainly because the amount of gas integrated along most
lines-of-sight is small, the features being very narrow. At
$\psi=22.5\arcdeg$, when the line-of-sight is nearly parallel to the
shock loci, they 
contribute the two straight and parallel segments separating the intense and
faint regions of the PVD just inside ${\mbox D}=\pm 2$. The high density
enhancements near the ends of the bar major axis contribute the very strong
linear segment going through the center of the PVDs at $\psi=0\arcdeg$ and
22.5\arcdeg. For these angles, they are seen roughly as segments of
circles. For viewing angles near 90\arcdeg, they give rise to the undulating
parts of the PVDs at large radius.
\placefigure{fig:nuclear}

We can repeat the masking process to keep only the region of the simulation
containing the nuclear spiral, as shown in Figure~\ref{fig:nuclear}. Comparing
the PVDs thus obtained with those of the $x_2$ family of periodic orbits
(Fig.~5 of \markcite{ba99}Paper~I), we find again many similarities, but also
some differences. In both cases, we see either an inverted S-shaped
feature or a near 
straight segment passing through the center of the PVDs. We refer the reader to
\markcite{ba99}Paper~I for a discussion and explanation of these shapes. More
similarity is found, however, if we compare PVDs at viewing angles differing
by $\approx20\arcdeg$, e.g.\ comparing the gaseous PVD at $\psi=112.5\arcdeg$
to the orbital one at $\psi=90\arcdeg$. This offset seems to corresponds to an
offset of the nuclear spiral with respect to the minor axis of the bar (and to
the offset of the straight shocks with respect to the major axis, see
\markcite{a92b}A92b), although this is hard to measure in the surface density
plot. Also, the PVDs of the nuclear spiral are more asymmetric with respect to
the viewing angles 0\arcdeg\ or 90\arcdeg\ than the PVDs of the outer bar
region (e.g.\ comparing the PVDs at $\psi=22.5\arcdeg$ and
$\psi=-22.5\arcdeg=157.5\arcdeg$). This is due to the weaker symmetry of the
nuclear spiral.

Because the streamlines in the central region are elongated roughly
perpendicular to the bar, like the $x_2$ orbits, higher radial velocities
(compared to the circular velocity) are reached when the bar is seen side-on
than when the bar is seen end-on. Considering the maximum velocity of the PVDs
as a function of the viewing angle, we find that it is highest for
$\psi=112.5\arcdeg$ and 135\arcdeg\ and lowest for $\psi=0\arcdeg$ and
22.5\arcdeg. The ``hole'' in the parallelogram-shaped signature of the $x_2$
orbits (see Fig.~5 of \markcite{ba99}Paper~I) has completely disappeared in the
hydrodynamical simulation, because the $x_2$-like behaviour of
the gas flow persists past the inner ILR and the flow becomes almost circular
in the very center.
\placefigure{fig:outer}

Figure~\ref{fig:outer} isolates the signature of the outer parts of the
simulation in the PVDs. Because the influence of the bar decreases rapidly
with radius, the flow outside the bar is close to circular. A perfectly
circular orbit would yield an identical inclined straight line passing through
the origin in all PVDs. The structure seen here can thus be thought of as a
succession of near-circular orbits of increasing radii, yielding the ``bow
tie'' signature observed. The ``hole'' seen in the center of the PVDs at
certain viewing angles (e.g.\ $\psi=45\arcdeg$) is again due to the fact that
the orbits are not perfectly circular. The strong almost solid-body features
forming loop-like structures near the upper and lower limits of the envelope
of the signature are due to tightly wound spiral arms in the outer disk.
\subsection{Model 086 (no-ILRs)\label{sec:086}}
\nopagebreak
\placefigure{fig:086}

Figure~\ref{fig:086} shows the face-on surface density distribution of model
086, which has no ILRs (or, equivalently, has no $x_2$ periodic orbits), and
the PVDs obtained using an edge-on projection. The main difference with the
density distribution of model 001 is the absence of any significant nuclear
spiral. The strong straight and narrow features in the center of the bar are
centered shock loci, caused by the high curvature of the streamlines near the
major axis of the bar (see Fig.~4 in \markcite{a92b}A92b). Similarly, the
$x_1$ orbits in this model have high curvatures or loops near their major axes
(\markcite{a92a}A92a). The shock loci do not curve near the center because no
streamline perpendicular to the bar exists; there are no $x_2$ periodic orbits
in this model. The shocks, however, still drive an inflow of gas, resulting in
the {\em entire} bar region being gas depleted. As expected, the region
outside the bar is almost unaffected by the change in the bar axial ratio
compared to model 001.

The PVDs for model 086 reflect all of the above properties. In particular,
they lack the central feature associated with the nuclear spiral in model 001,
and the signature of the bar region is very faint (Fig.~\ref{fig:086} should
be contrasted with Fig.~\ref{fig:001}). Because of the absence of an
$x_2$-like flow in the center of the simulation, the $x_1$-like flow extends
all the way to the center. In good agreement with what happens for the
periodic orbits (see \markcite{ba99}Paper~I or \markcite{a92a}A92a), the
streamlines are much more eccentric than those of model 001. The
parallelogram-shaped envelope of the signature of the bar region in the PVDs
therefore reaches even more extreme radial velocities (compared to the
circular velocity) than in model 001, and the rising part of the envelope
extends almost all the way to the center for viewing angles close to 0\arcdeg\ 
(see also Fig.~9 in \markcite{ba99}Paper~I). When the bar is seen close to
end-on, radial velocities more than twice those in the outer parts are
reached. These velocities decrease rapidly as the viewing angle approaches
$\psi=90\arcdeg$. Unfortunately, the signature of the bar region in the
PVDs is very faint, because of the strong inflow of gas toward the center. As
in the case of model 001, the regions outside the bar lead to a strong almost
solid-body component in all PVDs.
\subsection{Other Models\label{sec:others}}
\nopagebreak
In this section, we will analyse sequences of simulations corresponding to the
ranges along the axes of parameter space likely to be occupied by real
galaxies. We hope thereby to further our understanding of the PVDs of edge-on
barred spiral galaxies, and to be able to extend the bar diagnostics which we
will develop. We also wish to develop criteria allowing us to constrain the
bar properties and mass distribution of edge-on systems. To achieve this, we
will concentrate on understanding the PVDs of the gas flow within the barred
region of the simulations, and will use extensively for that purpose the
results of \markcite{a92a}A92a and \markcite{a92b}A92b. In particular,
\markcite{a92b}A92b showed that nuclear spirals occur in models with an
extended $x_2$ family of periodic orbits and lead to offset shocks, while
centered shocks occur in models with no or shortly extended $x_2$ orbits. She
also showed that $x_1$ periodic orbits with a high curvature near the major
axis of the bar are essential to the formation of shocks, and that such shocks
lead to an inflow of gas toward the central regions of the simulations,
depleting the outer (or entire) bar regions.
\placefigure{fig:a/b}

Figure~\ref{fig:a/b} shows a sequence of simulations with varying bar axial
ratio $a/b$ (the other parameters are kept fixed at those of model 001). The
figure shows, for each simulation, the face-on surface density distribution
and the velocity field (with a few streamlines) in the frame of reference
corotating with the bar. It also shows the PVDs obtained for various viewing
angles with respect to the bar when the simulation is viewed edge-on. To limit
the size of the figure, we show a slightly reduced number of viewing angles
compared to Figures~\ref{fig:001}--\ref{fig:086}.

For small bar axial ratios, when the shocks are short and curved, the outer
bar region is not strongly gas depleted and its signature is easily visible in
the PVDs. As we consider simulations with increasingly high bar axial ratios,
the density of the region of the PVDs which corresponds to the outer bar
region drops considerably. It reaches its lowest values for the highest bar
axial ratio considered ($a/b=5.0$), for which the shock loci are straight and
close to the bar major axis. The envelope of the signature of the outer bar
region in the PVDs becomes more extreme as the bar axial ratio is increased:
the maximum radial velocities reached for small viewing angles increase and
the positions of the maxima get closer to the center. These behaviours are
quasi-linear and the opposite is true for large viewing angles. For example,
at $\psi=0\arcdeg$, the maximum velocity of the outer bar region is about 250
for $a/b=1.5$, 360 for $a/b=3.0$, and 430 for $a/b=5.0$. At $\psi=90\arcdeg$,
the opposite is observed, with maximum velocities of about 225 and 215,
respectively. This is easily understood because, as the bar axial ratio
increases, the eccentricity of the streamlines in the outer bar region (and
that of the $x_1$ orbits, see \markcite{a92a}A92a) also increases, leading to
higher velocities along the line-of-sight when the bar is seen end-on, and
lower velocities when the bar is seen side-on. The most important effect of an
increase of the bar axial ratio is the disappearance of the
nuclear spiral for axial ratios $a/b\gtrsim2.7$. Because the nuclear spiral is
associated with an $x_2$-like flow, and because the range of radii occupied
by the $x_2$ orbits decreases rapidly as the bar axial ratio is increased (see
Fig.~6 of \markcite{a92a}A92a), the inverted S-shaped signature of the nuclear
spiral in the PVDs disappears for large bar axial ratios. The maximum radial
velocity reached by the nuclear spiral signature varies
little with the bar axial ratio (when present).
\placefigure{fig:rl}

Figure~\ref{fig:rl} shows how the PVDs of the simulated disks change when the
Lagrangian radius of the mass model is varied. \markcite{a92b}A92b showed that
the gas flow in a bar has shock loci offset toward the leading sides of the
bar and of the form observed in early-type barred spiral galaxies only for a
restricted range of Lagrangian radii, namely $r_L=(1.2\pm0.2)a$. All the
observational estimates for early-type strongly barred spirals also give
values within this range (see, e.g., \markcite{a92b}A92b; Elmegreen
\markcite{e96}1996). We will thus concentrate on this range here. Model 028,
the limiting case with $r_L=5.0$, shows strong spiral arms starting at the
ends of the bar and extending to large radii. The spiral arms are easily
identified in the PVDs as long filamentary structures. This model has no
nuclear spiral, and thus no corresponding inverted S-shaped feature in the
PVDs. The opposite is true for models with larger Lagrangian radii or,
equivalently, lower pattern speeds. Those models have extended $x_2$ families
of periodic orbits and therefore nuclear spirals (see \markcite{a92b}A92b). As
the pattern speed of the bar is further decreased, the radial range occupied
by the $x_2$ periodic orbits is increased and the nuclear spiral becomes more
predominant. The outer bar region decreases accordingly (\markcite{a92a}A92a).
This effect is clearly seen in the PVDs of Figure~\ref{fig:rl}. For increasing
Lagrangian radii, they show an increase of the radial extent of the nuclear
spiral signature, and a decrease of the radial extent of the outer bar region
signature. There is also an increase of the maximum radial velocity reached by
the nuclear spiral, but the effect is rather small.
\placefigure{fig:rhoc}

Figure~\ref{fig:rhoc} shows the behaviour of the PVDs as the central
concentration of the mass model is varied. For low central concentrations, the
gas streamlines are oval-shaped and aligned with the bar, there are no shocks,
and the bar region is only slightly gas depleted. The signature of the bar
region in the PVDs is then clear. As the central concentration is increased,
the streamlines become more eccentric and the envelope of the signature of
the bar region extends to higher velocities. Once the central concentration
reaches $\rho_c\approx2.2\times10^4$, an $x_2$-like flow appears in the center
of the bar and a nuclear spiral and offset shocks are formed.  These changes
are also easily seen in the PVDs, which acquire an inverted S-shaped feature,
while the density in the outer bar region drops. The region occupied by the
$x_2$ orbits then increases with increasing central concentration
(\markcite{a92a}A92a), and so does the radial extent of the nuclear spiral
signature in the PVDs. An increase in the central concentration of the mass
model has similar effects to an increase of the Lagrangian radius (see
Fig.~\ref{fig:rl}). This is not surprising, since both changes influence the
location of the resonances in a similar way.
\placefigure{fig:qm}

Figure~\ref{fig:qm} shows a sequence of simulations with varying bar
quadrupole moment. For the lowest quadrupole moment, the bar is weak and the
flow is close to circular, with only a weak nuclear spiral and curved shocks
in the center. This does not cause substantial inflow, and the gas density in
the barred region remains high. The velocity field shows that there is a
transition from a $x_1$-like flow to a $x_2$-like flow in the central region,
but the effect is not strong since the eccentricity of the streamlines is
small. All these effects reflect themselves in the PVDs. For somewhat higher
bar quadrupole moments (model 058), the nuclear spiral is well-developed and
its signature is strong in the PVDs, with a gap present between it and the
solid-body signature of the outer parts of the galaxy. As the bar quadrupole
moment is increased further, the envelope of the signature of the outer bar
region in the PVDs becomes more extreme, reaching larger radial velocities
and extending closer to the center. For $Q_m\gtrsim5.5\times10^4$, the nuclear
spiral disappears. Those two effects are due to the facts that the
eccentricity of the streamlines increases significantly with increasing bar
quadrupole moment, while the region occupied by the $x_2$ orbits decreases
until it disappears completely (see \markcite{a92a}A92a). Because the bar is
so strong for high quadrupole moments, the flow is non-circular even in the
outer parts of the simulations, and a ``hole'' is present in the center of the
PVDs at intermediate viewing angles. Not surprisingly, the variations in the
gas distribution and kinematics are similar when the bar quadrupole moment is
increased and when the axial ratio of the bar is increased.
\section{Dust Extinction\label{sec:dust}}
\nopagebreak
It is interesting to note that the PVDs discussed so far have a certain degree
of symmetry with respect to the viewing angles 0\arcdeg\ or 90\arcdeg. This
means that although it is relatively easy to determine whether a line-of-sight
is close to the major or the minor axis of a bar ($|\psi|$), it is considerably
harder to determine in which half of the PVD (positive or negative projected
distances from the center) the near side of the bar is located ($\pm\psi$).

This situation can be contrasted to that in the Galaxy, where most studies
have no difficulty identifying the quadrant in which the near side of the
Galactic bar is located. Studies using infrared photometry (e.g.\ Dwek et al.\
\markcite{detal95}1995; Binney, Gerhard, \& Spergel \markcite{bgs97}1997),
star counts (e.g.\ Weinberg \markcite{w92}1992; Stanek et al.\
\markcite{suskkmk97}1997), gaseous or stellar kinematics (e.g.\ Binney et al.\
\markcite{bgsbu91}1991; Wada et al.\ \markcite{wthh94}1994; Zhao, Spergel, \&
Rich \markcite{zsr94}1994; see also Beaulieu \markcite{b96}1996; Sevenster et
al.\ \markcite{ssvf97}1997), and microlensing events (e.g.\ Paczynski et al.\
\markcite{psuskkmk94}1994) all indicate a bar making an angle of 15\arcdeg\ to
45\arcdeg\ with respect to the line-of-sight to the Galactic center (positive
values indicating that the near side of the bar is at positive Galactic
longitude). In the case of the Galaxy, at least two effects help the observer
determine the exact orientation of the bar. Firstly, projection effects (both
in longitude and latitude) mean two lines-of-sight on each side of the
Galactic center reach correspondingly different parts of the bar (e.g.\ Binney
et al.\ \markcite{bgsbu91}1991). Secondly, the large difference between the
distances to each side of the bar means point sources in the far side of the
bar will appear significantly fainter than the corresponding sources in the
near side (e.g.\ Stanek et al.\ \markcite{suskkmk97}1997). In addition, the
far side of the bar will appear thinner that the near side (Dwek et al.\
\markcite{detal95}1995). For a galaxy at infinity, all lines-of-sight are
parallel, and no projection or distance effects are present. However, extinction within an
edge-on disk can play a similar role to that of distance in the Galaxy, and
can help constrain the orientation of a bar. Because the velocity spread in
the inner parts of the PVDs is so large, it is unlikely that self-absorption
by any line would be significant. Given the prominence of the dust lane in
many edge-on spiral galaxies, extinction by dust is likely to be the dominant
factor affecting the PVDs. If dust is present in significant amount, the
spectroscopic signature of the far side of the bar in a PVD should be fainter
than that of the nearer side.

\placefigure{fig:dust}

Figure~\ref{fig:dust} shows the surface density and PVDs of model 001 when
considering a dust distribution proportional to the gas surface density.
Because our simulations are not self-consistent, only the relative values of
the density are important and the value of the dust absorption coefficient per
unit mass $\kappa$ we use is meaningless. Our goal in this section being
merely to illustrate the effects of dust on the simulated PVDs, and not to
reproduce quantitatively the situation in real galaxies, we have simply
increased the value of $\kappa$ (in which the dust to gas ratio is also
folded) until the PVDs were significantly affected.
In Figure~\ref{fig:dust}, 
we have decreased the surface densities in the surface density plot 
to reflect the effective contribution of each point to
the projected density for a viewing angle $\psi=45\arcdeg$. Unlike
Figure~\ref{fig:001}, the PVDs are now far from symmetric with respect to
viewing angles of 0\arcdeg\ or 90\arcdeg. In addition, PVDs at intermediate
viewing angles are no longer antisymmetric with respect to the center. This is
mainly because the nuclear spiral obscures most of the material behind it,
breaking the symmetry of the parallelogram-shaped signature of the $x_1$-like
flow in the outer bar region. For viewing angles
$0\arcdeg\lesssim\psi\lesssim90\arcdeg$, the nuclear spiral obscures mostly
material moving away from the observer in the outer bar region, and obscures
only a small amount of material moving toward the observer in the same region.
Thus, the signature of the $x_1$-like flow in the PVDs is weakened for
positive radial velocities, leading to a much fainter signature of the outer
bar region in the upper halves of the PVDs than in the lower halves. The
opposite is true for viewing angles $90\arcdeg\lesssim\psi\lesssim180\arcdeg$,
where the signature of the outer bar region is much fainter in the lower
halves. This effect is strongest and least extended for viewing angles
$\psi\approx110\arcdeg$ to 135\arcdeg, as the line-of-sight is then roughly
parallel to the major axis of the nuclear spiral (which is slightly offset
from that of the $x_2$ periodic orbits). Some effects on the signature of the
nuclear spiral itself and on the signature of the outer parts of the
simulation are present, but they are less pronounced.

In addition to the diagnostics suggested in the previous sections to identify
a bar in an edge-on spiral galaxy and determine whether it is seen end-on or
side-on, the introduction of dust in the simulations has allowed us to develop
criteria to determine in which half of the galaxy the near side of the bar is
located. Of course, the distribution of dust in real galaxies will inevitably
be more complex than the highly idealised distribution adopted
here. Nevertheless, the features due to dust in the PVDs of
Figure~\ref{fig:dust} can probably still be used as a guide to interpret
asymmetries present in real data.
\section{Discussion\label{sec:discussion}}
\nopagebreak
\subsection{Bar Diagnostics\label{sec:diag}}
\nopagebreak
Unlike periodic orbits studies (\markcite{km95}KM95; \markcite{ba99}Paper~I),
where one has to adopt a way of populating the orbits, hydrodynamical
simulations provide both the velocity and density of the gas. In particular,
if some periodic orbit families intersect, it is not necessary to make a
choice between them, because the simulations will reveal which family the gas
follows in each region (and to what extent).  Nevertheless, as we will discuss
later in this section, the comparison with observations can still present some
problems, since observations involve the strength of a given emission line
rather than the gas density.
 
Notwithstanding the problem of populating the orbits, there is generally a
good agreement between the PVDs obtained from periodic orbit calculations
in \markcite{ba99}Paper~I and those obtained here from hydrodynamical
simulations. This is because there are several similarities between the
periodic orbits structure of the models and the gas flow
(\markcite{a92b}A92b). We have in both cases a central PVD component, which we
identified in \markcite{ba99}Paper~I with the $x_2$ orbits and in this paper
with the nuclear spiral. Further out, we have the domain of the $x_1$ orbits,
which covers the outer (or entire) bar regions of the simulations. In
\markcite{ba99}Paper~I, by studying in detail the signatures of various
periodic orbit families in the PVDs, we gained useful insight into the
projected structure and kinematics of the gas from first principles. This
allowed us to obtain a deeper understanding of the PVDs produced in the
hydrodynamical simulations.
 
The main feature of the PVDs is the gap (fainter region), present at all
viewing angles, between the signature of the nuclear spiral and that of the
outer parts of the simulations. Such a structure would not be possible in an
axisymmetric galaxy and it unmistakably reveals the presence of a bar or oval
in an edge-on disk. This gap occurs because of the large-scale shocks which
are present in bars, and which drive an inflow of gas toward the center,
depleting the outer bar regions.

The gaps present in the PVDs produced with periodic orbits (see
\markcite{km95}KM95; \markcite{ba99}Paper~I) are different in nature from the
ones observed here, being mainly due to the absence of populated orbits in
certain regions of the models, particularly near corotation. As a result, in
these studies, the low density region extends well beyond corotation, although
its exact extent depends on which orbits are neglected (e.g.\ as
self-intersecting) and how the other orbits are populated.  For example, in
Figure~1 of \markcite{km95}KM95, the gap extends to almost twice the
corotation radius.

We recall that 3D $N$-body simulations of rotating disks produce bars which,
when viewed edge-on, appear boxy-shaped if seen end-on and peanut-shaped if
seen side-on (see, e.g., Combes \& Sanders \markcite{cs81}1981; Combes et al.\ 
\markcite{cdfp90}1990; Raha et al.\ \markcite{rsjk91}1991). Taking the maximum
height of the peanut-shaped bars to occur at half the corotation radius
(Combes et al.\ \markcite{cdfp90}1990), we find that, in the
\markcite{km95}KM95 
case, the ratio of the radial extent of the gap in the PVDs to the radius
where the maximum height of the peanut-shaped bulges occurs should be
approximately 4 (or slightly less because of projection effects, associating
the bulges with edge-on bars). On the other hand, in the hydrodynamical
simulations presented here, the low density region of the PVDs reflects the
low density region in the outer parts of the bar, and should thus be within
corotation. When a bar is seen side-on, the ratio of the extent of the gap to
the radius of maximum height of the peanut-shaped bulge should thus be
approximately 2 (or slightly less). \markcite{km95}KM95 report that this ratio
is about 2 for the two galaxies they studied, in agreement with the prediction
of our hydrodynamical simulations. The same can be said for the galaxies
studied by Bureau \& Freeman (\markcite{bf97}1997) and \markcite{bf99}BF99.
These results can only be reconciled with the periodic orbits approach if the
maximum height of the peanut-shaped bulges occurs near corotation (rather than
halfway).

As we pointed out in \S~\ref{sec:001}, the shapes of the features present in
the PVDs vary with the viewing angle, and can thus in principle be used to
constrain that quantity in an observed galaxy. The envelope of the signature
of the outer bar region in the PVDs reaches very high radial velocities
(compared to those in the outer parts of the simulations) for viewing angles
close to the bar, and relatively low velocities for viewing angles
perpendicular to it.  However, the envelope is so faint in most cases that it
is unlikely to be of much use with real data (see \S~\ref{sec:others}). The
signature of the nuclear spiral has an inverted S-shape for some viewing
angles and is almost solid-body for others. That feature is rather thick and
does not show much fine structure, however, so it may be hard to use in
conjunction with observations. The best viewing angle diagnostic is probably
provided by the ratio of the maximum radial velocity reached by the nuclear
spiral component to the velocity in the outer parts of a galaxy. This ratio
should be greater than unity for viewing angles roughly perpendicular to the
bar and smaller than unity for viewing angles parallel to it.  Analysing their
sample, \markcite{bf99}BF99 find that the above is in good agreement with
their results, provided they make the reasonable assumption that bars are
peanut-shaped when seen close to side-on and boxy-shaped when seen close to
end-on (see, e.g., Combes \& Sanders \markcite{cs81}1981; Combes et al.\ 
\markcite{cdfp90}1990). Other properties of the PVDs, such as the maximum
radial velocity reached or the faintness of the signature of the outer (or
entire) bar regions, can also help to constrain the mass distribution and bar
properties of an observed galaxy (see \S~\ref{sec:others}). In practice,
however, the analysis and modeling of spectroscopic data should not be a
trivial task.

If a barred system has no nuclear spiral (or, equivalently, does not have an
extended $x_2$ family of periodic orbits), there will not be a nuclear spiral
signature in the PVDs and no gap, or double-peaked structure. In addition, the
surface density in the bar region can be extremely low. An example of such a
model was discussed in \S~\ref{sec:086} (model 086). In such cases, the only
component likely to be detected observationally is the solid-body signature of
the outer parts of the galaxy, and the kinematical detection of the bar will
not be straightforward. The first step would be to rule out such a slowly
rising rotation curve (the rotation curve being defined as the upper limit of
the envelope of an observed PVD), for example by calculating the shape of the
rotation curve expected from surface photometry. Unfortunately, this would
only show that the type of gas observed (ionised, neutral, or molecular) is
not present in large quantities in the central region of the galaxy, but it
would not provide any information on the cause of this depletion. It is thus
probably necessary to use stellar kinematics to identify a bar in such
systems. Because a large percentage of the stars in the central regions of
barred disks are expected to be trapped around the $x_1$ orbits, there should
be a clear bar signature in the PVDs. We shall explore possible diagnostics
based on the stellar kinematics (and corresponding PVDs) in
\markcite{ab99}Paper~III.

Cases with no nuclear spiral component should be a minority, however, at least
among strongly barred early-type spiral galaxies, since observational evidence
argue that these systems possess ILRs (see, e.g., Athanassoula
\markcite{a91}1991; \markcite{a92b}A92b). In particular, the fact that the
dust lanes in most early-type strongly barred galaxies are offset along the
leading sides of the bar, rather than centered and close to its major axis, is
a strong argument for the existence of ILRs (\markcite{a92b}A92b). In
addition, out of 17 galaxies with a boxy/peanut-shaped bulge in the sample of
\markcite{bf99}BF99, only four have no nuclear component in their PVD. The
lack of a nuclear component in these four galaxies could be due either to the
lack of ILRs, or to the lack of emitting gas around the ILRs region. At least
13 out of 17 galaxies, i.e. an overwhelming majority, have ILRs. Nevertheless,
the existence of ILRs deserves further study, particularly for later type
spirals. This can be done, for example, by high resolution kinematical
studies, to show the change of direction of the orbits or streamline
ellipticities in the central parts of galaxies (e.g.\ Teuben et al.\ 
\markcite{tsaa86}1986), or by calculating the families of periodic orbits in
galaxy potentials derived from observations, to show the existence or
non-existence of the $x_2$ (and $x_3$) family.
\subsection{NGC~5746-Like PVDs\label{sec:ngc5746}}
\nopagebreak
Although our simulations cover a fair fraction of the available parameter
space (see Table~1 in \markcite{a92a}A92a), we have never come across a PVD
showing a clear ``figure-of-eight'', as suggested by \markcite{km95}KM95. In
other words, the upper envelope of the low density region of the simulated
PVDs is never as pronounced as it is in NGC~5746 or, to a lesser extent, in
NGC~5965 and NGC~6722 (see \markcite{km95}KM95; Bureau \& Freeman
\markcite{bf97}1997; \markcite{bf99}BF99).

Based on periodic orbit calculations, we expect this region of the PVDs to
originate from material near the ends of the bar on its {\em minor} axis (see,
e.g., Fig.~2b in \markcite{ba99}Paper~I). Some models, particularly $n=0$
models, indeed have secondary density enhancements near the edge of the bar,
just outside the offset shocks which we have already discussed (see Fig.~3 in
\markcite{a92b}A92b). These enhancements are due to the small distance between
the outer ILR and the tip of the $x_1$ family characteristic curve in the
characteristic diagram (\markcite{a92b}A92b). This leads to orbit crowding, 
particularly near the bar minor axis. In these models, the upper envelope of
the gap in the PVDs is stronger than that in most other models.
\placefigure{fig:088}

Figure~\ref{fig:088} shows two density distributions for model 088 and the
corresponding PVDs for a
viewing angle $\psi=45\arcdeg$. On the left (Fig.~\ref{fig:088}a,b), the
entire density distribution and PVD are displayed. On the right
(Fig.~\ref{fig:088}c,d), most of the {\em PVD} was masked out, leaving only 
the region of interest in objects like NGC~5746. The corresponding density
distribution was obtained using a simple inversion scheme. The strong
upper envelope of the low density region of the PVD is clearly due to the
secondary density enhancement on the leading side of the bar. A similar effect
is observed in many $n=0$ models with low bar axial ratio $a/b$, small
Lagrangian radius $r_L$, and/or low central concentration $\rho_c$ (e.g.\ 
models 013-016, 040-043, 061-063, 087-088, and 108-110). Although the
intensities obtained in these simulations are much lower than the strong
envelopes observed in objects like NGC~5746, we must remember that it is the
density which is plotted here, and the gaseous emission in the secondary
density enhancements could be somehow enhanced, e.g.\ through shocks. Features
similar to those density enhancements are observed in real barred galaxies,
where ``plumes'' are sometime seen on the leading sides of bars (in NGC~1365
for example). We suggest here that these plumes may be at the origin of the
strong upper envelope of the gap observed in the PVD of some galaxies.

We must stress, however, that a strong envelope is unusual; it is absent from
the PVD of most galaxies. In the sample of \markcite{bf99}BF99, only 2
galaxies out of 17 present clear indications of such a feature. This seems in
agreement with the explanation presented above, as strong ``plumes'' are not
often observed.
\subsection{Limitations of the Models\label{sec:limits}}
\nopagebreak
When using the bar diagnostics developed in this paper to interpret
observational data, one has to take into account the fact that the PVDs were
calculated using the gas {\em density} in the simulations and not the strength
of a given gaseous emission line, which is what one usually observes. Physical
conditions in the gas vary across the simulated disks, and will lead to
different excitation mechanisms dominating in different parts of the
disks. Regions of high density and low shear 
are likely to have higher star formation rates and thus more
photoionised gas than elsewhere. Conversely, the gas in or near shocks, such
as the nuclear spiral, will be mainly shock-excited (see Binette, Dopita, \&
Tuohy \markcite{bdt85}1985; Dopita \& Sutherland \markcite{ds96}1996), with
possible star formation depending on the shear. Because different excitation
mechanisms lead to different emission line ratios, the relative amplitude of
the various components of a PVD (e.g.\ the intensity of the signature of the
outer parts of the disk versus that of the nuclear spiral) will depend on the
emission line used in the observations. For example, Bureau \& Freeman
(\markcite{bf97}1997) and \markcite{bf99}BF99 found that, for many objects,
the signature of the nuclear spiral was very strong in the
[\ion{N}{2}]~$\lambda$6548,6584 lines but was almost absent in H$\alpha$, 
probably indicating that it is shock-excited (the other components of
the PVDs showed [\ion{N}{2}]~$\lambda$6584/H$\alpha$ ratios typical of
\ion{H}{2} regions). The PVDs were not corrected for stellar absorption,
however, so these results should be interpreted with caution. Nevertheless,
the PVDs produced should only be used as guide when interpreting kinematical
data, and one should have a basic understanding of the mechanisms involved in
the production of a given line before using the PVDs for comparison
purposes. We stress that the morphology of the PVDs (the multiple components)
is a more significant bar diagnostic than the distribution of intensity (the
relative amplitude of the components).

Conversely, under certain assumptions about the physical conditions in the
gas, it is possible to use the observed emission line ratios (e.g.\ 
H$\alpha$/H$\beta$) to measure the extinction due to dust in data.
This could prove useful to interpret asymmetries present in observed PVDs (see
\S~\ref{sec:dust} for a more complete discussion of likely dust effects).

Contrary to the ``building blocks'' approach of \markcite{ba99}Paper~I, which
used combinations of periodic orbit families to model the structure and
kinematics of barred galaxies, the hydrodynamical simulations presented here
inherently take into account the collisional nature of gas, so the kinematics
of the gaseous component is more accurately modeled. Approximations to the gas
properties are nevertheless necessary, and we recall that the gas was treated
as ideal, isothermal, infinitely thin, and non-self-gravitating. How much do
our results depend on these assumptions, or, in other words, how model
dependent are they?
 
The interstellar medium is a complicated multi-phase mixture, which can only
be described schematically, particularly in non-local studies covering an
object the size of a galaxy. Two different approaches have been developed so
far. In the first one, the gas is treated as ballistic particles which, when
they collide, lose energy according to pre-specified recipes (e.g.\ Miller,
Prendergast, \& Quirk \markcite{mpq70}1970; Schwarz \markcite{s81}1981,
\markcite{s84}1984; Combes \& Gerin \markcite{cg85}1985). Unfortunately, the
results of these simulations can depend on the adopted collision law (Guivarch
\& Athanassoula \markcite{ga99}1999), at least as far as the shocks in the
barred region are concerned. In the second approach, a hydrodynamical
treatment is used, solving the Euler equations with the help of a grid
(e.g.\ van Albada \& Roberts \markcite{vr81}1981; van Albada, van Leer, \&
Roberts \markcite{vvr82}1982; Mulder \markcite{86}1986; Piner, Stone, \&
Teuben \markcite{pst95}1995), Smooth Particles Hydrodynamics (SPH; Lucy
\markcite{l77}1977; Gingold \& Monaghan \markcite{gm77}1977; Hernquist \& Katz
\markcite{hk89}1989; etc), or beams (Sanders \& Prendergast
\markcite{sp74}1974). A number of these studies assume an isothermal equation
of state, following the model of Cowie \markcite{c80}(1980), who calculated
the equation of state for an ensemble of clouds and found it could be
described as isothermal, provided the clouds have an equilibrium mass
spectrum.

Comparing the results of all these schemes is beyond the scope of this paper.
In all approaches, the more reliable codes produce shocks in the bar region
which are more or less offset from the bar major axis towards its leading
sides. These shocks result in an inflow of gas and a substantial lowering of
the density in the outer bar region. Although the precise location, shape, and
persistence of the shocks differ, those are relatively small effects, compared
to the fact that we weight by density rather than by the strength of an
emission line. The main reason for adopting the present hydrodynamical code is
that it gave very good results in many previous studies (see
\markcite{a92b}A92b).
 
Because dissipation ensures that the gas layer in spiral galaxies remains
thin, the two-dimensional nature of the simulations should not be a factor
limiting the applicability of the results to the interpretation of real data.
Furthermore, vertical motions have no direct consequence on the PVDs produced
as the movement is perpendicular to the line-of-sight. It is somewhat harder 
to gauge the consequences of the fact that we have
ignored the gas self-gravity. Because our mass model represents best (barred)
early-type spirals (see \markcite{a92a}A92a), where the fraction of the total
mass in gas is typically less than 10\%, the contribution of the gas to the
potential and large scale forces is negligible. Indeed, Lindblad, Lindblad, \&
Athanassoula (\markcite{lla96}1996) found that including self-gravity in their
model of NGC~1365 made very little difference to the global picture. It may
well be, however, that near high density structures such as the nuclear
spiral, the self-gravity of the gas is important.
\section{Summary and Conclusions\label{sec:conclusions}}
\nopagebreak
Our main aim in this paper, the second in a series, was to develop diagnostics
to identify bars in edge-on spiral galaxies using the particular kinematics of
the gaseous component of barred disks. To achieve this goal, we ran
two-dimensional hydrodynamical simulations of the gas flow in the potential of
a barred spiral galaxy mass model. We constructed position-velocity diagrams
(PVDs) from those simulations, using an edge-on projection and various viewing
angles with respect to the major axis of the bar. The presence of shocks and
inflows in the simulations allowed us to develop better bar diagnostics than
those presented in our first paper (Bureau \& Athanassoula
\markcite{ba99}1999), based on periodic orbits calculations.

We analysed in detail two simulations which are prototypes of simulations for
models with and without inner Lindblad resonances (which we associate with the
existence of $x_2$ periodic orbits). We showed that, for models allowing $x_2$
orbits, the nuclear spiral which is created in the center of the simulations
produces a strong inverted S-shaped signature in the PVDs. This signature
reaches high radial velocities (compared to those in the outer parts of the
simulations) when the bar is seen side-on, and relatively low velocities when
the bar is seen end-on. The flow in the outer bar region (the entire bar
region if no nuclear spiral is present) produces a parallelogram-shaped
signature in the PVDs, being associated with the $x_1$ periodic
orbits. Because the flow is mostly along the bar in that region, the
highest velocities are now reached for viewing angles close to the bar major
axis. In the outer parts of the simulations, the flow is almost circular and
produces a strong almost solid-body signature in the PVDs for all viewing
angles.

Shocks within the bar are present in most simulations and lead to a depletion
of the gas in the region of the bar occupied by an $x_1$-like flow. Thus, if a
nuclear spiral is present, a bar can easily be identified in an edge-on spiral
galaxy, as there will be a gap in the PVD between the signature of the nuclear
spiral and that of the outer parts of the galaxy. If there is no nuclear
spiral, it may still be possible to detect a bar, but only with the help of
photometry and/or stellar kinematics. The envelope of the signature of the
nuclear spiral, and to a lesser extent that of the outer bar region, is most
useful to determine the orientation of a bar with respect to the
line-of-sight. It is nevertheless hard to discriminate between two viewing
angles on either side of the bar. We showed that adding dust to the
simulations helps break this degeneracy.

We also produced PVDs for a range of simulations covering most of the fraction
of parameter space likely to be occupied by real galaxies. These simulations
can be used to constrain the mass distribution and bar properties of an
observed system. In particular, the presence or absence of the signature of a
nuclear spiral in a PVD places strong constraints on the values the parameters
of our mass model may take. The nuclear spiral can be absent for high bar
axial ratios and/or bar quadrupole moments, and for low Lagrangian radii
and/or central concentrations.
\acknowledgments
We thank K.\ C.\ Freeman, A.\ Bosma, and A.\ Kalnajs for comments on the
manuscript and J.-C.\ Lambert for his computer assistance. E.\
A.\ thanks G.\ D.\ Van Albada for making available to her his version
of the FS2 code. M.\ B.\ acknowledges the support of an Australian
DEETYA Overseas 
Postgraduate Research Scholarship and a Canadian NSERC Postgraduate
Scholarship during the conduct of this research. M.\ B.\ would also like to
thank the Observatoire de Marseille for its hospitality and support during a
large part of this project.

\clearpage
%
% Figure captions
%
% Figure containing the density and PVDs for model 001.
%
\figcaption{Gas distribution and position-velocity diagrams of model 001, with
  $n=1$, $a/b=2.5$, $r_L=6.0$, $\rho_c=2.4\times10^4$, and
  $Q_m=4.5\times10^4$. The upper-left plot shows the face-on logarithmic gas
  surface density of the simulation. We use logarithmic values (in this plot
  alone) because the dynamic range of the gas surface density is high. Dark
  shades denote low density regions. The bar is at 45\arcdeg\ to the
  horizontal, has a semi-major axis length of 5~kpc, and is rotating
  clockwise. The other plots show position-velocity diagrams (projected
  density of material as a function of line-of-sight velocity and projected
  position along the major axis) of the gas when the simulation is viewed
  edge-on. For those plots, dark shades denote high density regions. The angle
  between the line-of-sight and the bar is indicated in the top-left corner of
  each diagram, a viewing angle of $\psi=0\arcdeg$ indicating that the bar is
  seen end-on (line-of-sight parallel to the bar) and a viewing angle of
  $\psi=90\arcdeg$ indicating that the bar is seen side-on (line-of-sight
  perpendicular to the bar). The viewing angle is illustrated in the surface
  density plot and increases counterclockwise.
\label{fig:001}}
%
% Figure containing the density and PVDs of the outer bar region for model
% 001.
%
\figcaption{Same as Figure~\ref{fig:001} but for the low density outer bar
  region of model 001 only.
\label{fig:bar}}
%
% Figure containing the density and PVDs of the nuclear spiral region for model
% 001.
%
\figcaption{Same as Figure~\ref{fig:001} but for the high density nuclear
  spiral region of model 001 only.
\label{fig:nuclear}}
%
% Figure containing the density and PVDs of the outer regions for model
% 001.
%
\figcaption{Same as Figure~\ref{fig:001} but for the high density outer
  regions of model 001 only.
\label{fig:outer}}
%
% Figure containing the density and PVDs of model 086.
%
\figcaption{Same as Figure~\ref{fig:001} but for model 086, with $n=1$,
  $a/b=5.0$, $r_L=6.0$, $\rho_c=2.4\times10^4$, and $Q_m=4.5\times10^4$.
\label{fig:086}}
%
%\pagebreak
%
% Figure containing the density and PVDs for the bar axial ratio sequence.
%
\figcaption{Gas distribution and position-velocity diagrams for a sequence of
  models of increasing bar axial ratio $a/b$. The models are 010, 005, 001,
  083, and 086, with, respectively, $a/b=$ 1.5, 2.0, 2.5, 3.0, and 5.0. The
  other parameters of the models are kept fixed at those of model 001: $n=1$,
  $r_L=6.0$, $\rho_c=2.4\times10^4$, and $Q_m=4.5\times10^4$. The first row of
  plots shows the face-on logarithmic gas surface density of the simulations.
  We use logarithmic values (in these plots alone) because the dynamic range of
  the gas surface density is high. Dark shades denote low density regions. The
  bar is at 45\arcdeg\ to the horizontal, has a semi-major axis length of
  5~kpc, and is rotating clockwise. The second row of plots shows velocity
  vectors and gas streamlines in the frame of reference corotating with the
  bar. The following rows of plots show position-velocity diagrams (projected
  density of material as a function of line-of-sight velocity and projected
  position along the major axis) of the gas when the simulations are viewed
  edge-on. For those plots, dark shades denote high density regions. The angle
  between the line-of-sight and the bar is indicated in the top-left corner of
  the first diagram of each row, a viewing angle of $\psi=0\arcdeg$ indicating
  that the bar is seen end-on (line-of-sight parallel to the bar) and a
  viewing angle of $\psi=90\arcdeg$ indicating that the bar is seen side-on
  (line-of-sight perpendicular to the bar). The viewing angle increases
  counterclockwise in the surface density plots.
\label{fig:a/b}}
%
% Figure containing the density and PVDs for the Lagrangian radius sequence.
%
\figcaption{Same as Figure~\ref{fig:a/b} but for a sequence of models of
  increasing Lagrangian radius $r_L$. The models are 028, 001, and 030, with,
  respectively, $r_L=$ 5.0, 6.0, and 7.0. The other parameters of the models
  are kept fixed at those of model 001: $n=1$, $a/b=2.5$,
  $\rho_c=2.4\times10^4$, and $Q_m=4.5\times10^4$.
\label{fig:rl}}
%
% Figure containing the density and PVDs for the central density sequence.
%
\figcaption{Same as Figure~\ref{fig:a/b} but for a sequence of models of
  increasing central density $\rho_c$. The models are 080, 066, 001, 069, and
  150, with, respectively, $\rho_c=$ $0.4\times10^4$, $1.6\times10^4$,
  $2.4\times10^4$, $3.2\times10^4$, and $4.0\times10^4$. The other parameters
  of the models are kept fixed at those of model 001: $n=1$, $a/b=2.5$,
  $r_L=6.0$, and $Q_m=4.5\times10^4$.
\label{fig:rhoc}}
%
%\pagebreak
%
% Figure containing the density and PVDs for the bar quadrupole moment sequence.
%
\figcaption{Same as Figure~\ref{fig:a/b} but for a sequence of models of
  increasing bar quadrupole moment $Q_m$. The models are 138, 058, 001, 052,
  and 120, with, respectively, $Q_m=$ $0.5\times10^4$, $2.5\times10^4$,
  $4.5\times10^4$, $6.0\times10^4$, and $8.0\times10^4$. The other parameters
  of the models are kept fixed at those of model 001: $n=1$, $a/b=2.5$,
  $r_L=6.0$, and $\rho_c=2.4\times10^4$.
\label{fig:qm}}
%
% Figure containing the density and PVDs of model 001 with dust.
%
\figcaption{Same as Figure~\ref{fig:001} but considering a dust distribution
  proportional to the gas surface density. The absorption coefficient per unit
  mass $\kappa=0.1$. The surface densities in the top-left plot have been
  decreased to reflect the effective contribution of each point to the
  integrated density along the line-of-sight for a viewing angle
  $\psi=45\arcdeg$. New features in the PVD for $\psi=45\arcdeg$ can thus be
  directly related to new features in the surface density plot just above it.
\label{fig:dust}}
%
% Figure containing the density and PVDs of model 088, with inversion scheme.
%
\figcaption{Gas distributions and position-velocity diagrams of model 088,
  with $n=0$, $a/b=3.0$, $r_L=5.5$, $\rho_c=2.4\times10^4$, and
  $Q_m=4.5\times10^4$. (a) shows the face-on logarithmic gas surface density
  of the simulation. We use logarithmic values (in the density plots alone)
  because the dynamic range of the gas surface density is high. Dark shades
  denote low density regions. The bar is at 45\arcdeg\ to the horizontal, has
  a semi-major axis length of 5~kpc, and is rotating clockwise. (b) shows the
  position-velocity diagram (projected density of material as a function of
  line-of-sight velocity and projected position along the major axis) of the
  gas when the simulation is viewed edge-on. For this plot, dark shades
  denote high density regions. The angle between the line-of-sight and the bar
  is $\psi=45\arcdeg$, a viewing angle of $\psi=0\arcdeg$ indicating that the
  bar is seen end-on (line-of-sight parallel to the bar) and a viewing angle
  of $\psi=90\arcdeg$ indicating that the bar is seen side-on (line-of-sight
  perpendicular to the bar). The viewing angle increases counterclockwise in
  the surface density plots. (c)--(d) are the same as (a)--(b) except
  that most of the simulation has been masked out. Only the upper envelope of
  the low density region of the {\em PVD} has been kept, as well as the
  corresponding region of the density distribution, obtained using a simple
  ``inversion'' scheme.
\label{fig:088}}
\end{document}